%% file: ANA-STDM-2019-01-PAPER.tex
\documentclass[cernpreprint, atlasdraft=false, texlive=2023, UKenglish, orcidlogo, texmf, orcidlogo]{atlasdoc}
 
\usepackage{atlaspackage}
\usepackage{atlasbiblatex}
 
\usepackage{atlasphysics}
\usepackage{atlasprocess}

\graphicspath{{logos/}{figures/}}
 
\usepackage{ANA-STDM-2019-01-PAPER-defs}

\AtlasTitle{Measurement of the Z boson invisible width at $\sqrt{s}=13$ TeV with the ATLAS detector}
 
\AtlasVersion{2.1}
 
\AtlasAbstract{
A measurement of the invisible width of the $Z$ boson using events with jets and missing transverse momentum is presented using 37 $\mbox{fb\(^{-1}\)}$ of 13 TeV proton--proton data collected by the ATLAS detector in 2015 and 2016. The ratio of $Z\rightarrow \textrm{inv}$ to $Z\rightarrow\ell\ell$ events, where inv refers to non-detected particles and $\ell$ is either an electron or a muon, is measured and corrected for detector effects. Events with at least one energetic central jet with $p_{\textrm{T}} \geq 110$ GeV are selected for both the $Z\rightarrow \textrm{inv}$ and $Z\rightarrow\ell\ell$ final states to obtain a similar phase space in the ratio. The invisible width is measured to be $506\pm2 \textrm{ (stat.)} \pm12 \textrm{ (syst.)}$ MeV and is the single most precise recoil-based measurement. The result is in agreement with the most precise determination from LEP and the Standard Model prediction based on three neutrino generations.
}

\AtlasRefCode{STDM-2019-01}
 
\PreprintIdNumber{CERN-EP-2023-232}

\AtlasJournal{Phys.\ Lett.\ B.}
\AtlasJournalRef{Phys. Lett. B 854 (2024) 138705}
\AtlasDOI{10.1016/j.physletb.2024.138705}

\hypersetup{pdftitle={ATLAS document},pdfauthor={The ATLAS Collaboration}}
 
\begin{document}
 
\maketitle

\section{Introduction}

Precision measurements of fundamental parameters of the Standard Model, such as properties of the \ZBoson, are crucial probes of the Standard Model and may provide hints of new physics. A measurement of the invisible width of the \ZBoson, \ZinvGamma, reveals the number of light neutrinos that couple to the \ZBoson~and any potential non-Standard Model contributions. Measurements of \ZinvGamma via different final states and analysis strategies test the consistency of the Standard Model  as differences amongst measurements would indicate possible sources of new physics.
 
The most precise measurement of the invisible width of the \ZBoson\ is that obtained by the LEP experiments via measurements of the total \ZBoson\ width~\cite{bib:lepzinvindaleph,bib:lepzinvinddelphi,bib:lepzinvindl3,bib:lepzinvindopal} and subtracting the partial width to visible final states: pairs of leptons (electrons, muons, taus) and hadrons. The invisible width is measured to a precision of 0.3\% assuming charged lepton universality, and 0.5\% without this assumption~\cite{ALEPH:2005ab}. Measurements of \ZinvGamma using photon-tagged events were also performed by the LEP experiments with a combined precision of 3.2\%~\cite{bib:lepzinvdiropal,bib:lepzinvdirl3,bib:lepzinvdiraleph,PDG2022}. The CMS Collaboration measured \ZinvGamma using a jet+$\MET$ final state, where \MET is the missing transverse momentum, with a precision of 3.2\% dominated by systematic uncertainties in the electron and muon identification and jet energy scale~\cite{CMS-SMP-18-014}. Jet+$\MET$ final states are also sensitive to anomalous production of missing transverse momentum with dedicated analyses generally targeting regions with large \MET~\cite{EXOT-2018-06,CMS-EXO-20-004}.
 
This paper presents a measurement of \ZinvGamma using the ratio of \Zinvj to \Zllj cross sections, defined as
\begin{eqnarray}
\notag	\Rmiss (\Zpt) & \equiv & \left( \frac{d\sigma(\Zinvj)}{d\Zpt} \right) / \left( \frac{d\sigma(\Zllj)}{d\Zpt} \right) \\
& = & \left( \frac{d\sigma(\Zjets)\times BR(\Zinv)}{d\Zpt} \right) / \left( \frac{d\sigma(\Zjets)\times BR(\Zboson\rightarrow\ell\ell)}{d\Zpt} \right),
\label{eqn:ratio}
\end{eqnarray}
 
where $\ell$ refers to electrons and muons, \Zpt is the \ZBoson\ \pt, as defined in Section~\ref{sec:selection} and $BR(\Zboson\rightarrow\ell\ell)$ are the leptonic branching ratios measured by the LEP experiments. In this ratio, the numerator and the denominator are corrected to a common phase space so that many of the dominant systematic uncertainties cancel. After all corrections, the value of the ratio is constant as a function of \Zpt. \ZinvGamma is determined by utilising the measured value of the leptonic width~\cite{ALEPH:2005ab, PDG2022} and fitting a constant value of $\Rmiss (\Zpt)$:
\begin{equation}
\ZinvGamma = \Rmisshat \cdot \ZllGamma,
\label{eqn:ratioGamma}
\end{equation}
 
where $\ZllGamma$ is the leptonic width and $\Rmisshat$ is the result of the fit.
 
The \Zinv, \Zees, and \Zmms\ regions are defined to require at least one energetic jet. This criterion is applied to all regions to ensure that the dominant systematic uncertainties in the individual processes cancel in the ratio. Residual differences in the phase space between the numerator and the denominator of Equation~\ref{eqn:ratio} are accounted for using simulations.

\section{ATLAS detector}
 
The ATLAS experiment~\cite{PERF-2007-01} at the LHC is a multipurpose particle detector with a forward-backward symmetric cylindrical geometry and a near \(4\pi\) coverage in solid angle\footnote{ATLAS uses a right-handed coordinate system with its origin at the nominal interaction point (IP) in the centre of the detector and the $z$-axis along the beam pipe. The $x$-axis points from the IP to the centre of the LHC ring, and the $y$-axis points upward. Cylindrical coordinates $(r,\phi)$ are used in the transverse plane, $\phi$ being the azimuthal angle around the $z$-axis. The pseudorapidity is defined in terms of the polar angle $\theta$ as $\eta=-\ln\tan(\theta/2)$.}. It consists of an inner tracking detector surrounded by a thin superconducting solenoid providing a \SI{2}{\tesla} axial magnetic field, electromagnetic and hadron calorimeters, and a muon spectrometer. The inner tracking detector covers the pseudorapidity range \(|\eta| < 2.5\). It consists of silicon pixel, silicon microstrip, and transition radiation tracking detectors. Lead/liquid-argon (LAr) sampling calorimeters provide electromagnetic (EM) energy measurements with high granularity. A steel/scintillator-tile hadron calorimeter covers the central pseudorapidity range (\(|\eta| < 1.7\)). The endcap and forward regions are instrumented with LAr calorimeters for both the EM and hadronic energy measurements up to \(|\eta| = 4.9\). The muon spectrometer surrounds the calorimeters and is based on three large superconducting air-core toroidal magnets with eight coils each. The field integral of the toroids ranges between \num{2.0} and \SI{6.0}{T\metre} across most of the detector. The muon spectrometer includes a system of precision tracking chambers and fast detectors for triggering. A two-level trigger system is used to select events. The first-level trigger is implemented in hardware and uses a subset of the detector information to accept events at a rate below \SI{100}{\kHz}. This is followed by a software-based trigger that reduces the accepted event rate to \SI{1}{\kHz} on average depending on the data-taking conditions. An extensive software suite~\cite{ATL-SOFT-PUB-2021-001} is used in the reconstruction and analysis of real and simulated data, in detector operations, and in the trigger and data acquisition systems of the experiment.

\section{Data set and simulated samples}
 
The data used were recorded with the ATLAS detector from 2015 to 2016 in $pp$ collisions at $\sqrt{s}=$ 13~\TeV\ and correspond to a total integrated luminosity of $36.65\pm0.32$~\ifb~\cite{DAPR-2021-01}. The mean number of $pp$ interactions per bunch crossing, including the hard scattering and other interactions in the same bunch crossings (pile-up), was $\langle \mu \rangle = 23$.
 
Monte Carlo (MC) simulation samples are used to correct the data for detector effects and to estimate contributions from some background events. The generated samples were processed using the \GEANT-based ATLAS detector simulation~\cite{SOFT-2010-01,Agostinelli:2002hh} and the same event-reconstruction algorithms are used for both the MC samples and the data.
 
The production of \ZBosons\ in association with jets for all leptonic decay modes was simulated with the ATLAS configuration of \Shtte~\cite{PMGR-2021-01}, which includes matrix elements for up to five partons at leading order (LO) and up to two partons at next-to-leading order (NLO). They are calculated with the \Comix~\cite{Gleisberg:2008fv} and \OpenLoops~\cite{Buccioni:2019sur,Cascioli:2011va,Denner:2016kdg} libraries and  matched with the \SHERPA parton shower (PS)~\cite{Schumann:2007mg} using the \MEandPSatNLO prescription~\cite{Hoeche:2011fd,Hoeche:2012yf,Catani:2001cc,Hoeche:2009rj} with a set of tuned parameters (`tune') developed by the \SHERPA authors. It includes a modified Catani--Seymour (CS) subtraction scheme~\cite{Catani:2002hc}, the Hessian \NNPDFthreeNNLO\ Parton Density Function (PDF) set~\cite{Ball:2014uwa}, and an analytic enhancement technique~\cite{PMGR-2021-01}. The prediction from \Shtte\ also includes NLO virtual Electroweak corrections~\cite{PMGR-2021-01} that were found to have a negligible effect on the measured width and are therefore not considered.
 
An alternative \Zjets\ MC signal sample, referred to as \MGfnlo~\cite{PMGR-2021-01}, was produced with the \MGNLO[2.6.5]~\cite{Alwall:2014hca} program to generate matrix elements at NLO accuracy in Quantum-Chromodynamics (QCD) for up to three additional partons in the final state. The \NNPDFthreeOneNNLO\ set~\cite{Ball:2014uwa} was used in the generation. The parton showering and subsequent hadronisation were performed using \PYTHIA[8.240]~\cite{Sjostrand:2014zea} with the A14 tune~\cite{ATL-PHYS-PUB-2014-021} and the \NNPDFtwoLO PDF set~\cite{Ball:2012cx}. The jet multiplicities were merged using the FxFx prescription~\cite{Frederix_2012}.
 
For background processes, \Wjets\ production is modelled with the same \SHERPA configuration as the nominal \Zjets\ production. Samples of diboson and triboson final states ($VV,~VVV$, where $V$ is either a \Wboson\ or \Zboson\ boson and indicates all possible multi-\Wboson\ or \Zboson\ boson combinations) were produced with the \Shtto\ or \Shttt\ generator depending on the process, including off-shell effects and Higgs boson contributions where appropriate. Fully leptonic final states and semileptonic final states, where one boson decays leptonically and the other hadronically, were simulated using matrix elements at NLO accuracy in QCD for up to one additional parton and at LO accuracy for up to three additional parton emissions.
The matrix element calculations were matched and merged with the \SHERPA parton shower using the \MEandPSatNLO prescription with a set of tuned parameters developed by the \SHERPA authors, and the \NNPDFthreeNNLO set of PDFs was used.
 
Electroweak production of \(\ell\ell jj\), \(\ell\nu jj\) and \(\nu\nu jj\) final states was simulated with \SHERPA[2.2.11]~\cite{Bothmann:2019yzt} using LO matrix elements with up to one additional parton emission. The matrix elements were merged with the \SHERPA parton shower~\cite{Schumann:2007mg} following the \MEPSatLO prescription~\cite{Catani:2001cc} and using the set of tuned parameters developed by the \SHERPA authors. The \NNPDF[3.0nnlo] set of PDFs~\cite{Ball:2014uwa} was employed. The samples were produced using the vector-boson-fusion approximation, which avoids overlap  with semileptonic diboson topologies by requiring a \(t\)-channel colour-singlet exchange. The starting conditions of the CS shower are set according to the large-\(N_c\) amplitudes supplied by Comix~\cite{Buckley:2021gfw} to achieve the correct vector-boson-fusion-appropriate radiation pattern.
 
Production of \ttbar\ is simulated at NLO accuracy in perturbative QCD using the \POWHEGBOX[v2]~\cite{Frixione:2007nw,Nason:2004rx,Frixione:2007vw,Alioli:2010xd} generator with the \NNPDFthreeNLO\ PDF set and the \hdamp\ parameter\footnote{The \hdamp\ parameter is a resummation damping factor and one of the parameters that control the matching of \powheg matrix elements to the parton shower and thus effectively regulates the high-\pt\ radiation against which the \ttbar\ system recoils.} set to 1.5~\mtop~\cite{ATL-PHYS-PUB-2016-020}. The events were interfaced to \PYTHIA8.230~\cite{Sjostrand:2014zea} to model the parton shower, hadronisation, and underlying event, with parameters set according to the A14 tune  and using the \NNPDFtwoLO\ set of PDFs. The \ttbar\ sample is normalised to the cross-section prediction at next-to-next-to-leading order (NNLO) accuracy, including the resummation of next-to-next-to-leading logarithmic (NNLL) soft-gluon terms calculated with \TOPpp[2.0]~\cite{Beneke:2011mq,Cacciari:2011hy,Baernreuther:2012ws,Czakon:2012zr,Czakon:2012pz,Czakon:2013goa,Czakon:2011xx}.
 
Single top quark production in the $s$-channel, in the $t$-channel, and in association with a $W$ boson ($tW$) was modelled using the \POWHEGBOX[v2] generator at NLO in QCD with the four-flavour scheme for the $t$-channel and five-flavour scheme for the $s$- and $tW$-channels and, the \NNPDFthreeNLO\ set of PDFs. The diagram-removal scheme~\cite{Frixione:2008yi} was used to remove interference and overlap with \ttbar\ production. The $tW$ cross section is corrected to the theory prediction at approximate NNLO accuracy~\cite{Kidonakis:2010ux,Kidonakis:2013zqa}, while the $s$- and $t$-channel cross sections are corrected to the prediction at NLO accuracy~\cite{Kant:2014oha-CUSTOM,Aliev:2010zk}.
 
For bottom and charm hadron decays, the \EVTGEN[1.7.0] program~\cite{EvtGen} was used for \MGfnlo\ samples, and \EVTGEN[1.2.0] was used for the \POWHEG samples. Pile-up was modelled by overlaying the simulated hard-scattering event with inelastic $pp$ events generated with \PYTHIA[8.186]~\cite{Sjostrand:2007gs} using the \NNPDFtwoLO\ PDF set and the A3 tune~\cite{ATL-PHYS-PUB-2016-017}. The small differences in lepton reconstruction, isolation, and trigger efficiencies between simulation and data are corrected in the simulation on an event-by-event basis by applying efficiency scale factors for each lepton~\cite{EGAM-2018-01,PERF-2015-10,MUON-2018-03}.

\section{Event selection}
\label{sec:selection}
 
This analysis uses only data taken during 2015--2016 to profit from lower \MET trigger thresholds. Events are used if they were recorded during stable beam conditions and if they satisfy detector and data-quality requirements~\cite{DAPR-2018-01}. They are required to have a primary vertex, defined as the vertex with the highest sum of track $\pt^2$, with at least two associated tracks with $\pt > \SI{500}{\MeV}$~\cite{PERF-2015-01}.  The triggers used to collect \Zinv and \Zmms events require an \MET threshold between 70--110 \GeV\ depending on the data-taking period~\cite{TRIG-2018-05,TRIG-2019-01,TRIG-2019-03}. In these trigger algorithms, input from the muon spectrometer is not included and therefore muons are treated like non-interacting particles. The trigger efficiencies plateau in the region of $\MET \ge 170~\GeV$. For the \Zees events, the data were collected using single-electron triggers, with \pT{} thresholds ranging from 20 to 26~\GeV{} and varying identification and isolation criteria.
 
Electron candidates are reconstructed from inner-detector tracks that come from the primary vertex and are matched to clusters of energy deposits in the EM calorimeter. They must satisfy likelihood-based identification requirements~\cite{EGAM-2018-01} based on EM shower shapes, track quality, and track--cluster matching. Baseline electron candidates are used if they pass the `Loose' identification requirements, and have $\pt \geq 7~\GeV$ and $|\eta|<1.37$ or $1.52 < |\eta| < 2.47$. For the signal region, electron candidates must in addition satisfy the `Tight' identification requirements and the `FCHighPtCaloOnly'~\cite{EGAM-2018-01} isolation requirement, which is based on calorimeter information only and has good background rejection in the high \pT region. To fulfil the primary-vertex condition, the signal-region electron track's transverse impact parameter significance must also satisfy $|d_0|/\sigma(d_0) < 5.0$, where $d_0$ is the transverse impact parameter and $\sigma(d_0)$ its uncertainty, and the longitudinal impact parameter $z_0$ must satisfy $|z_0\sin(\theta)| < 0.5$~mm, where $\theta$ is the angle of the track to the beamline.

Muon candidates are identified by matching inner-detector tracks from the primary vertex to either full tracks or track segments reconstructed in the muon spectrometer. They are required to satisfy identification requirements~\cite{PERF-2015-10,MUON-2018-03} based on quality criteria applied to the inner-detector and muon-spectrometer tracks. Baseline muon candidates are used in the analysis if they pass the `Loose' identification requirements and have $\pt \geq 7~\GeV$ and $|\eta|<2.5$. Signal-region muon candidates in addition are required to satisfy `Medium' identification requirements~\cite{PERF-2015-10,MUON-2018-03} and must satisfy the `FCLoose' isolation requirement, built from tracking and calorimeter information, with a muon-\pT-dependent variable cone size $\Delta R$~\cite{MUON-2018-03} for isolation to other tracks. The signal region candidates must satisfy the following primary-vertex requirements: the transverse impact parameter significance must satisfy $|d_0|/\sigma(d_0) < 3.0$ and the longitudinal impact parameter must satisfy $|z_0\sin(\theta)| < 0.5$~mm, where $d_0$, $\sigma(d_0)$, $z_0$ and $\theta$ are as defined above for the electrons.
 
Hadronically decaying taus are required to satisfy the `JETIDRNNLOOSE' selection~\cite{ATL-PHYS-PUB-2019-033}, based on a recurrent neural network using single-track variables, and reconstructed kinematic and topological variables. They must have either one or three tracks associated with the tau candidate with an absolute charge of 1. Tau candidates are used if they have $\pt > 20~\GeV$ and $|\eta|<1.37$ or $1.52 < |\eta| < 2.47$. Tau identification is used to reject background events.
 
Jets of hadrons are reconstructed using a particle-flow algorithm~\cite{PERF-2015-09} based on noise-suppressed positive-energy topological clusters in the calorimeter. Energy deposited in the calorimeter by charged particles is subtracted and replaced by the momenta of tracks that are matched to those topological clusters. The jets are clustered using the anti-\kt~\cite{Cacciari:2008gp} algorithm implemented in the \textsc{FastJet} package~\cite{Cacciari:2011ma} with a radius parameter $R = 0.4$. They are initially calibrated using simulations and corrected using in situ measurements~\cite{JETM-2018-05}. Jets are required to have a calibrated $\pT \geq 30~\GeV~$ and $|y |<4.4$. Jets with \pt~<~60~\GeV\ and $|\eta| <$ 2.5 need to satisfy the `Tight' jet vertex tagger (JVT) working point~\cite{ATLAS-CONF-2014-018}, which uses tracking information to suppress pile-up jets. The forward-jet-vertex tagger (fJVT)~\cite{ATL-PHYS-PUB-2019-026} identifies pile-up jets in the forward region with $|\eta| >$ 2.5 using the `Tight' working point that sets constraints on the timing and the momentum of the jet. Events in which a jet is removed by the JVT algorithms are vetoed to reduce contributions of topologies where a falsely removed jet leads to \MET.
To reduce calorimeter noise `Loose' jet quality criteria ~\cite{ATLAS-CONF-2015-029} are applied rejecting events with any jet potentially originating from anomalous energy depositions due to coherent noise and electronic noise bursts in the calorimeter~\cite{DAPR-2012-01}. For a further suppression of non-collision backgrounds either due to beam-induced or cosmic muons, the leading jet must satisfy the `Tight' cleaning criteria. This requires the ratio of the charge fraction, which is the scalar sum of the \pt of tracks over the scalar sum of the calorimeter energy in the jet, to the maximum sampling fraction of the jet energy collected by a single calorimeter layer to be larger than $0.1$.
 
Electrons, muons, taus, and jets are reconstructed and identified independently. An overlap-removal procedure is then applied to uniquely identify these objects in an event. Jets are removed if the $\Delta R$ between the jet and a lepton is smaller than 0.2. Then, electrons and muons closer than $\Delta R = 0.4$ to any remaining jet are removed.
 
The \MET is computed as the negative of the vectorial sum of the transverse momenta of tracks associated with electrons, muons, hadronically decaying taus and any other object classified as a jet. Tracks in the inner detector that are associated to the primary vertex but not with any other component are also included~\cite{PERF-2016-07}. A \Zpt variable is defined to have a common observable between all regions. For the \Zinv region, \Zpt is the \MET. For the \Zees and \Zmms regions, \Zpt is the vectorial sum of the \MET and the \pT of the two electrons or muons. For the \Wboson boson control regions, defined below, the \Zpt is the vectorial sum of the \MET and the \pt of the electron or muon.
 
All regions are required to have $\Zpt \ge 130~ \GeV$ and at least one jet with $\pT \geq 110~\GeV$ and $|\eta| < 2.4$. To suppress the multijet background, events are vetoed if the separation in azimuthal angle between any jet and the \Zpt\ vector, $\Delta \phi (\text{jet}, \Zpt)$,  is less than $0.4$. For the \Zinv region, events are vetoed if any baseline lepton, including any tau lepton, is present.
 
For the \Zees region, events are required to have exactly two electrons passing the `Tight' likelihood-based identification requirements, the `FCHighPtCaloOnly'~\cite{EGAM-2018-01} isolation requirement and the track-to-vertex association requirements. The leading (sub-leading) electron is selected with a \pt larger than $50\ (25)~\GeV$. This selection reduces background processes without removing many \Zees events as the \Zpt criterion selects \Zboson events where the leading lepton has a high \pT. The two electrons must have opposite charges and the invariant mass of the di-electron system needs to satisfy $66 < m_{ee} < 116~\GeV$. Events are vetoed if any additional baseline lepton, including any tau lepton, is present.
 
For the \Zmms region, events are required to have exactly two `Medium', isolated muons from the primary vertex with $\pt \geq 50\ (25)~\GeV$ for the leading (sub-leading) muon. The muons must have opposite charges and the invariant mass of the di-muon system must satisfy $66 < m_{\mu\mu} < 116~\GeV$. Events are vetoed if any additional baseline lepton, including any tau lepton, is present.
 
\section{Background estimates}
 
The dominant background processes in the \Zinv region are those from \Wjets production (38\% of the total number of events), multijet processes (5\%), top-quark production (6\%), diboson production, \Zlls production (less than 2\%) and non-collision events (1\%). Contributions from the \Wjets production are estimated using simulations but constrained by dedicated \Wjets control regions. The multijet and non-collision processes are estimated with data-driven methods. All other background processes are estimated using simulations. For the \Zees and \Zmms regions, the dominant background processes are top-quark production (2\% of the total number of events), diboson production (less than 2\%) and contributions from non-prompt and mis-reconstructed leptons (1\%). The latter is estimated via a data-driven method, whereas all others are determined via simulations.
 
\begin{figure}[h!]
\centering
\includegraphics[width=0.49\textwidth]{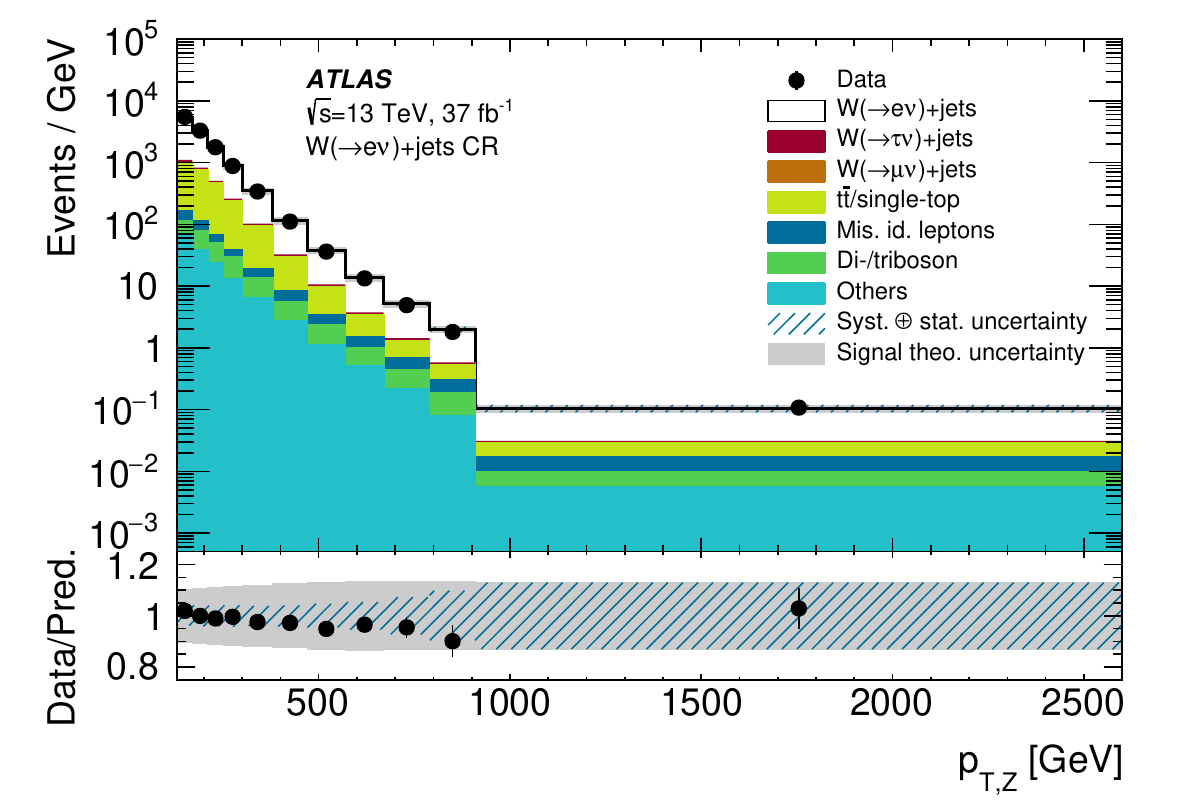}
\includegraphics[width=0.49\textwidth]{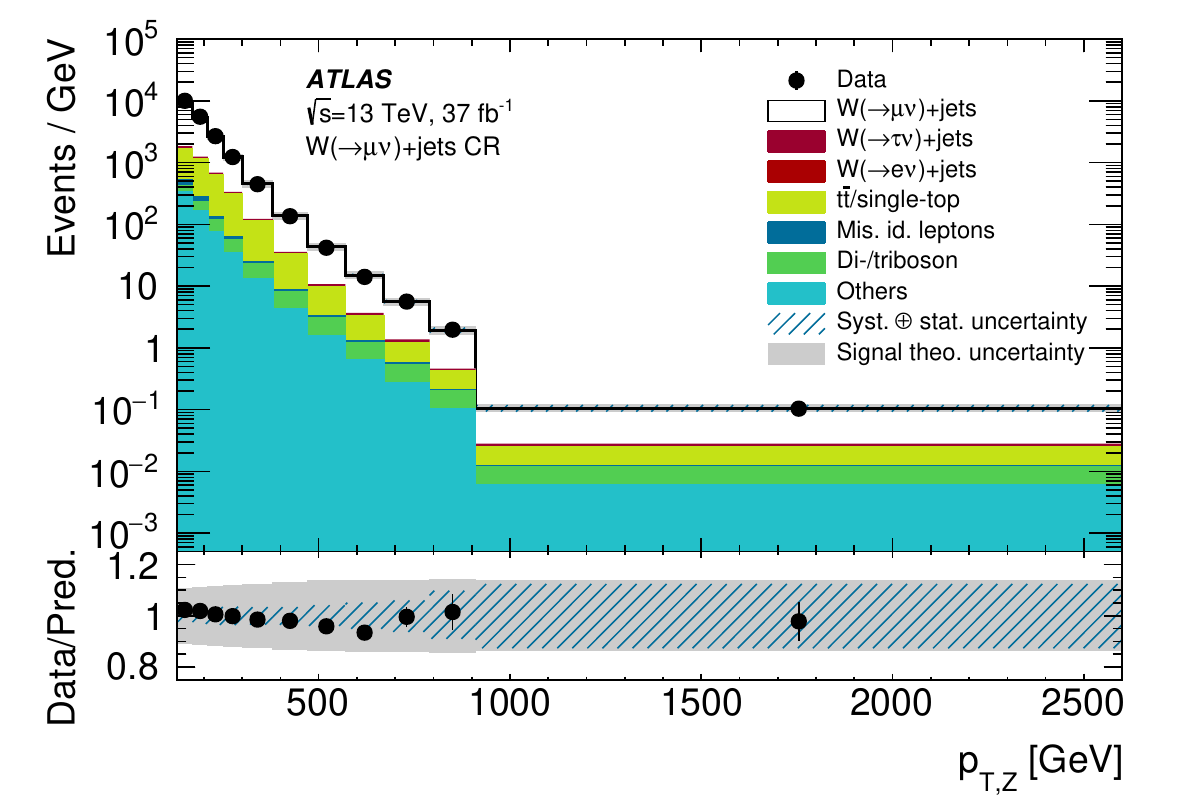}
\caption{Data-to-predictions comparisons in (left) \Wenj and (right) \Wmnj events as a function of \Zpt in the respective control regions. Statistical uncertainties in the data are shown as error bars and the total combined statistical and systematic uncertainties are shown as hashed bands. The theory uncertainties in the \Wens and \Wmns processes are shown by the light grey shaded bands. The experimental and theory uncertainties are shown separately and not added in quadrature.}
\label{fig:WlnuMet}
\end{figure}

The \Wens and \Wmns control regions are defined using the same \Zpt, jet and $\Delta \phi (\text{jet}, \Zpt)$ criteria as in the \Zinv region. The \Wens control region definition requires in addition one electron with the same characteristics as the leading electron in the \Zees region, $\MET > 50$ \GeV\ and a transverse mass $50 \leq \mT \leq 110~\GeV$, where $\mT = \sqrt{2\pt^\ell\MET (1-\cos{(\phi^\ell-\phi^{\MET})})}$, $\phi^\ell$ and $\phi^{\MET}$ are the azimuthal angles of the lepton and the \MET respectively. The \Wmns region requires exactly one muon as defined in the \Zmms region with \pt > 50 \GeV\ and $50 \leq m_{\text{T}} \leq 110~\GeV$. In both the control regions, backgrounds from non-prompt and mis-reconstructed leptons are estimated via the matrix method, as described below. All other backgrounds, which amount to less than 25\% of the total number of events, are estimated using simulations. Figure~\ref{fig:WlnuMet} shows the number of events as a function of \Zpt for the \Wens and \Wmns control regions.
 
The \Wens (\Wmns) simulations in the \Zinv region are scaled by the ratio of data to simulations obtained in the \Wens (\Wmns)-enhanced control region in each bin of the \Zpt distributions. The \Zpt-integrated scale factors for both channels are within 1\% of unity and agree with each other within their systematic uncertainties. For contributions from \Wtn, where obtaining a sufficiently pure control region is difficult, the \Wtn simulation is scaled by the average value of those obtained in the \Wens and \Wmns control regions. This average value was checked in a tau-enhanced region, defined by selecting hadronically decaying taus with a $\pt > 30 \GeV$. The average value was found to be appropriate in this region.
 
The multijet background in the \Zinv region originates largely from the mis-reconstruction of the energy of a jet in the calorimeter and, to a lesser extent, the presence of neutrinos in the final state from heavy-flavour hadron decays. The multijet background is determined from data, using the jet smearing method as described in Ref.~{\cite{SUSY-2011-20, EXOT-2018-06}}. It relies on the assumption that the \met value of multijet events is dominated by fluctuations in the jet response in the detector, which can be measured in the data. The method was validated using a region where events were selected as in the \Zinv region, except for a modified requirement that the minimum azimuthal distance between a jet and \MET is between 0.3 and 0.4, which indicates events with mis-measured jets. After the event selection, the multijet background is estimated to be about $8\%$ in the lowest \Zpt bin and falls below $1\%$ for \Zpt above 250~\GeV.
 
Non-collision backgrounds in the \Zinv region, which arise when beam-halo protons intercept the LHC collimators resulting in muon-producing particle cascades are estimated via the method outlined in Ref.~\cite{DAPR-2012-01}. In particular, the jet timing, $t_j$, calculated from the energy-weighted average of the time of the jet energy deposits, defined relative to the event time in nominal collisions, is used. A dedicated region enhanced in beam-induced background, defined by inverting the tight jet-quality selection imposed on the leading jet, is used to estimate the amount of non-collision background from the fraction of events with a leading-jet timing $|t_j| > \SI{5}{ns}$. The results indicate a contribution of 1.2\% from non-collision backgrounds in the signal region again steeply falling with \Zpt.
 
In the \Zees, \Zmms regions and the \Wjets control regions, contributions from non-prompt and mis-reconstructed electrons and muons are determined using the matrix method~\cite{ATLAS-CONF-2014-058}. In this approach, the number of events with a mis-identified electron or muon is determined via a set of linear equations, based on the number of events in the signal region, the number of events in a sample where the lepton selection criteria are loosened to ensure a high fraction of mis-identified leptons and the efficiencies of real and mis-identified leptons to satisfy the signal region lepton criteria. The sample with loosened lepton criteria is defined as events that satisfy the baseline lepton selection but fail to meet the additional signal region lepton criteria. The efficiency of real leptons to satisfy the signal region lepton criteria relative to the loosened lepton criteria is determined from MC samples. The efficiency for mis-identified leptons is determined in an independent control region. For the mis-identified electrons, this control region uses the same criteria as with the \Wens region but without applying the \Zpt and $\mT$ selection criteria and inverting the \MET criterion. For the mis-identified muons, the control region uses the same criteria as with the \Wmns region but without applying the $m_T$ selection criterion. With this method, the non-prompt and mis-reconstructed backgrounds are estimated to be about $1.0\%$ and $0.7\%$ in the \Zees and \Zmms regions, respectively. The background is roughly $1\%$ in the \Wjets control regions.
 
The matrix method is cross-checked for \Wens events using a data-driven template fit approach. Here, the \MET distribution for the mis-identified electron background is derived from data in a background-enriched control region. This region is defined by inverting the identification criteria and dropping the track-to-vertex association criteria. The contributions to the background template that do not originate from mis-identified electrons are evaluated and subtracted using simulation. A fit to the \MET distribution is performed to determine the number of events with non-prompt and mis-reconstructed electrons. The resulting fraction of mis-identified electron events in the signal region agrees within the respective uncertainties to that obtained from the matrix method.

Distributions of the \Zpt for the \Zinv, \Zees and \Zmms regions are shown in Figure~\ref{fig:detectlevel-srs}. The predictions agree well with the data within the experimental and theoretical uncertainties. The binning is optimised such that the bin width is at least two times the detector resolution. The highest \Zpt bin in each region has at least 20 events and the bin ends at 2600 \GeV\ above which the expected number of events is small.

\begin{figure*}[h!]
\centering
\includegraphics[width=0.51\textwidth]{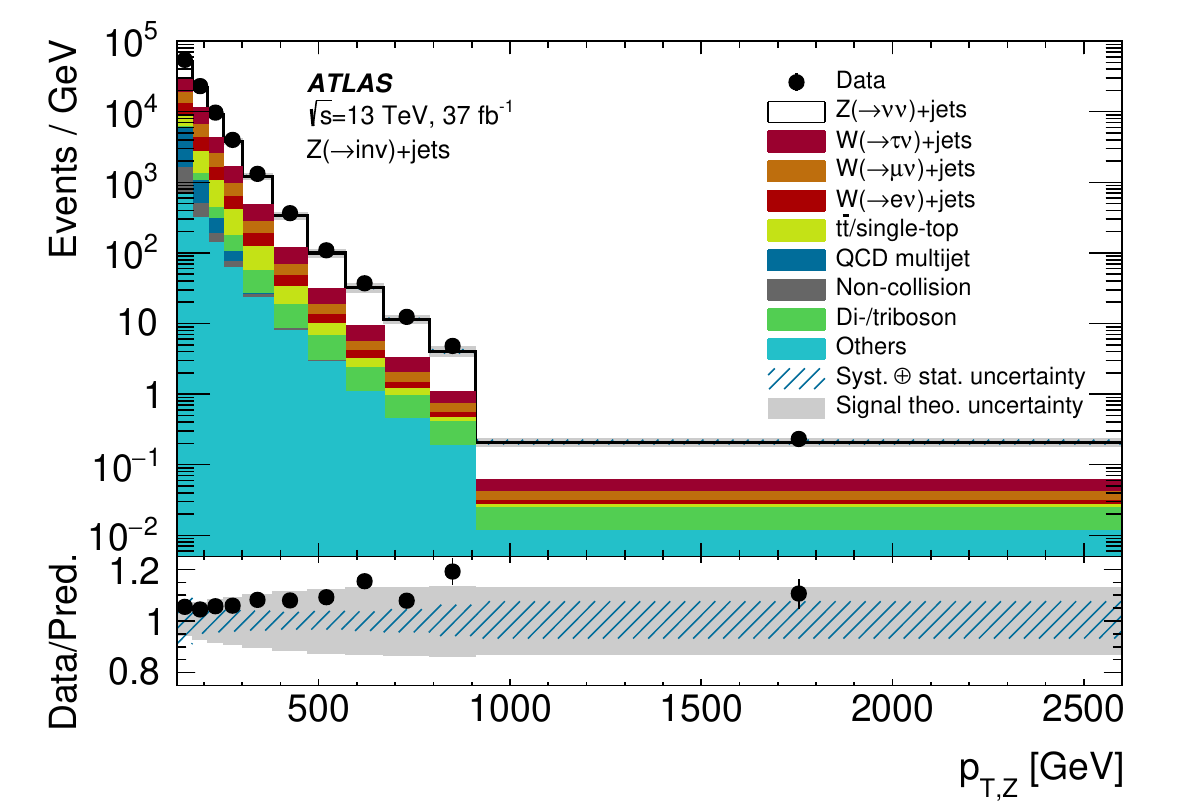}
\includegraphics[width=0.49\textwidth]{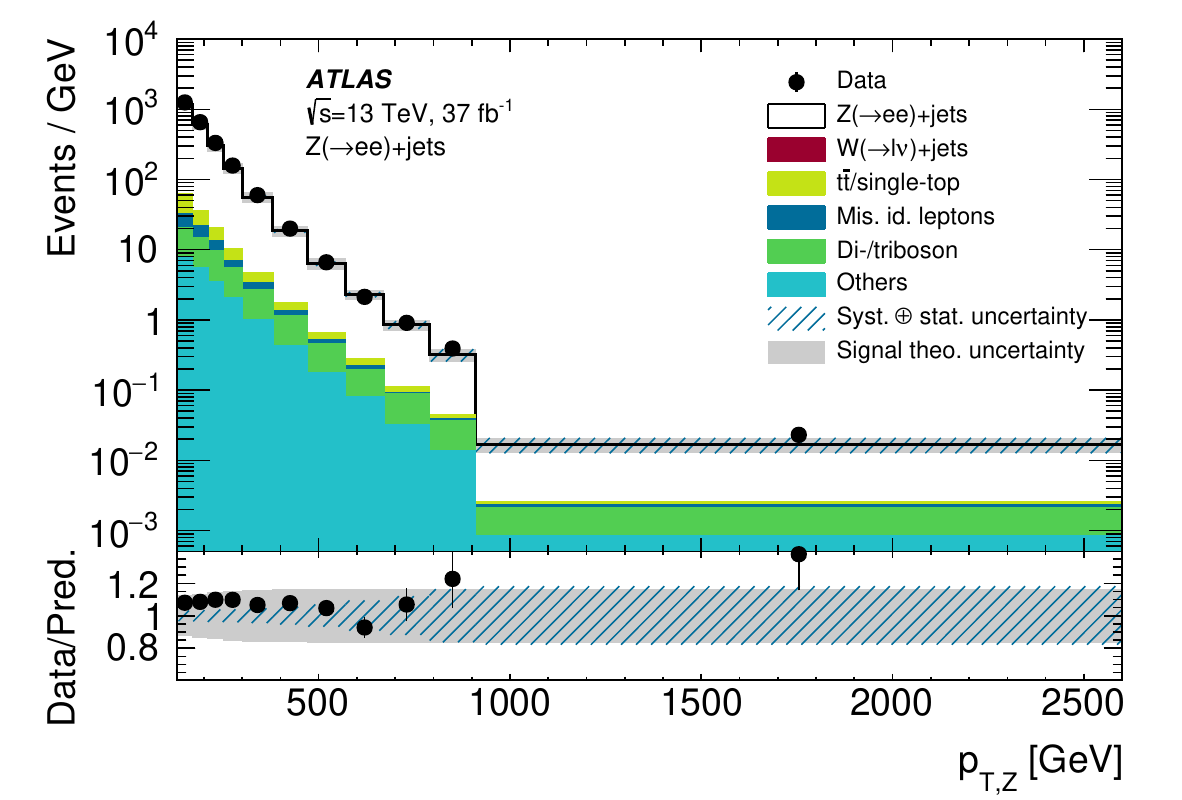}
\includegraphics[width=0.49\textwidth]{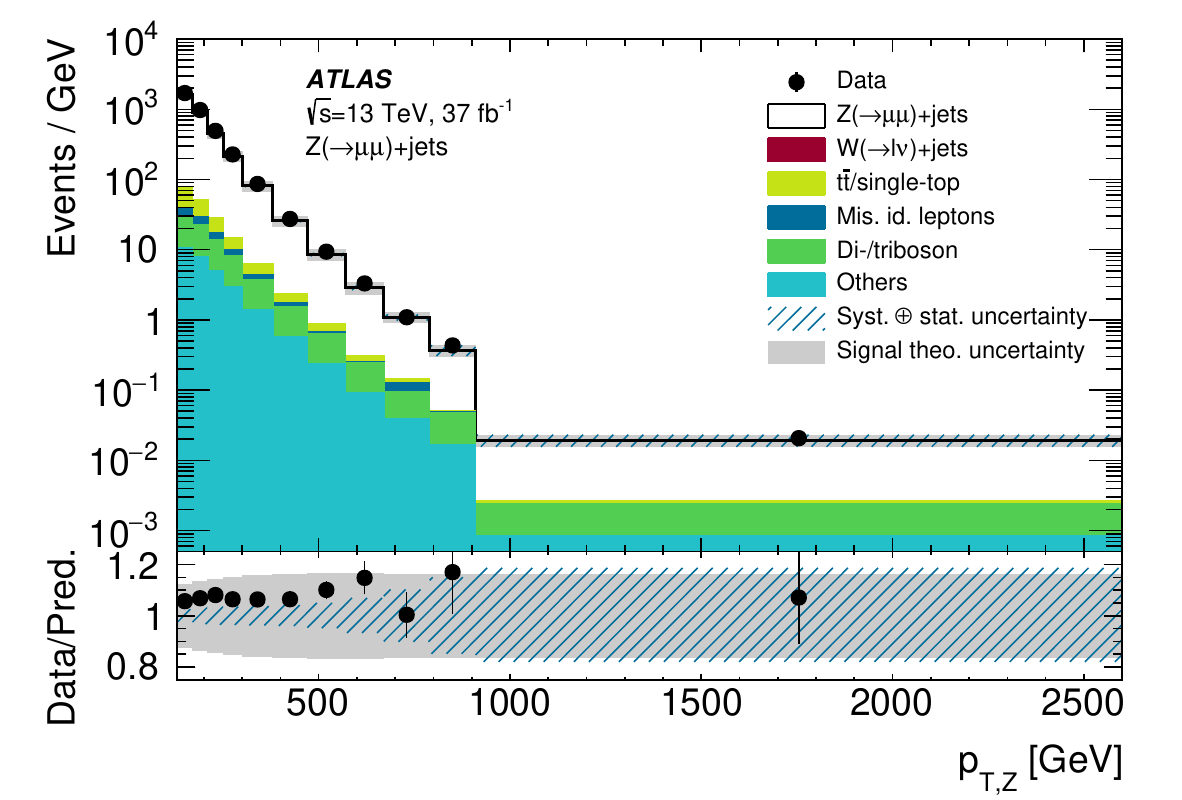}
\caption{Data-to-predictions comparisons in the \Zinv (top), \Zees (bottom left) and \Zmms (bottom right) signal regions as a function of \Zpt. In the \Zinv, the \WBoson\ scale factors are applied as described in the text. Statistical uncertainties in the data are shown as error bars and the total combined statistical and systematic uncertainties are shown as hashed bands. The theory uncertainties in the \Zinv, \Zees, and \Zmms processes are shown by the light grey shaded bands. The experimental and theory uncertainties are shown separately and not added in quadrature.}
\label{fig:detectlevel-srs}
\end{figure*}

\section{Systematic uncertainties}
 
Systematic uncertainties in the measured distributions and ratio stem from experimental and simulation modelling uncertainties. The uncertainties are estimated by applying them to simulation, and then re-estimating the background contributions and the detector-correction factors. The impact of the systematic uncertainties in the measured width is summarised in Table~\ref{tab:Uncertainties} and the relative uncertainties are shown in Figure~\ref{fig:unfUnc} in the appendix.
 
Systematic uncertainties associated with jet reconstruction are addressed via jet-energy-scale (JES) variations in a 29 nuisance-parameter scheme and jet-energy-resolution (JER) variations in a 13 nuisance-parameter scheme~\cite{JETM-2018-05}. The uncertainty in the efficiency to satisfy the (f)JVT requirement for pile-up suppression was derived in \Zmmj events and is also considered~\cite{PERF-2016-06}. The uncertainty in \met due to a possible miscalibration of its soft-track component was derived from data--simulation comparisons of the \pT balance between the hard and soft \met components~\cite{PERF-2016-07}. Imperfect modelling of the effects of pile-up leads to acceptance changes for different jet multiplicities. To assess this uncertainty, the average number of pile-up interactions is varied in simulation. The uncertainty in the combined 2015--2016 integrated luminosity is 0.87\%~\cite{DAPR-2021-01}, obtained using the LUCID-2 detector~\cite{LUCID2} for the primary luminosity measurements; this uncertainty cancels in the ratio, apart from small residual dependencies on the luminosity in the background estimates.
 
Systematic uncertainties in the lepton selection are related to the reconstruction, identification, isolation, and trigger~\cite{EGAM-2018-01,PERF-2015-10,MUON-2018-03} efficiencies and scale factors. Uncertainties in the electron energy scale and resolution are also taken into account~\cite{PERF-2017-01}. The largest impact on this result is that from the uncertainties on the electron identification and then to a lesser extent that from the uncertainties on the isolation. For the muons, the uncertainties are related to the muon momentum scale, inner-detector and muon-spectrometer resolution, and sagitta-bias correction~\cite{MUON-2022-01}. The largest impact due to the muon uncertainties are those from the isolation and reconstruction. For the $\tau$-leptons, energy scale uncertainties are included~\cite{ATLAS-CONF-2017-029}.
 
Modelling uncertainties in MC samples used in the signal and control regions are taken into account by varying the QCD scales and the PDF parameterisation. The effect of QCD scale uncertainties is defined by the envelope of variations resulting from changing the renormalisation (\muR) and factorisation (\muF) scales by factors of two with an additional constraint of $0.5\leq \muR / \muF \leq2$. Uncertainties due to the PDF parameterisation are evaluated using sets of PDF variations~\cite{Butterworth:2015oua} and consistent variations of \alphas\ in the PDF and in the hard scatter based on \NNPDFthreeNNLO~\cite{Ball:2014uwa}. The diagram-removal scheme~\cite{Frixione:2008yi} is used to remove interference and overlap between $tW$ and \ttbar\ production. A related uncertainty was estimated by comparing to an alternative sample simulated using the diagram-subtraction scheme~\cite{Frixione:2008yi,ATL-PHYS-PUB-2016-020}. The uncertainty in \ZllGamma is taken from the Particle Data Group combination~\cite{PDG2022}.
 
For the \Wens and \Wmns-enhanced control regions, these uncertainties are applied to all simulation-based contributions. For the non-prompt and mis-reconstructed electrons and muons control regions, only scale uncertainties in the simulation-based contributions are significant and therefore considered. For these control regions, an additional uncertainty is estimated by varying the selection criteria.  For the multijet background and non-collision background estimates, a conservative 100\% uncertainty is assigned.
 
To account for uncertainties on the detector correction, an uncertainty in the correction method itself is estimated by taking the difference between the corrected ratio using these bin-wise correction factors, as detailed below in Equation~(\ref{eqn:correctionFactor}), and when using two iterations of Bayesian unfolding~\cite{DAgostini:1995}.
 
\begin{table}[h!]
\centering
\begin{tabular}{l c c}
\toprule
Systematic Uncertainty \qquad Impact on \ZinvGamma & in [\MeV] & in [\%]\\
\midrule
Muon efficiency & 7.4 & 1.5\\
Renormalisation \& factorisation scales & 5.9 & 1.2\\
Electron efficiency & 4.9 & 1.0 \\
Detector correction & 4.4 & 0.9\\
QCD multijet & 3.2 & 0.6\\
\MET & 2.4 & 0.5\\
$Z(\rightarrow \mu\mu)+$jets misid. lepton estimate & 1.9 & 0.4\\
Jet energy resolution & 1.6 & 0.3\\
$W(\rightarrow \ell \nu)+$jets normalisation & 1.5 & 0.3\\
Pile-up reweighting & 1.5 & 0.3\\
Non-collision background estimate & 1.3 & 0.3\\
Jet energy scale & 1.3 & 0.3\\
$\gamma^*$-correction & 1.0 & 0.2\\
$Z(\rightarrow ee)+$jets misid. lepton estimate & 1.0 & 0.2\\
Luminosity & 1.0 & 0.2\\
Parton distribution functions + $\alpha_s$& 0.7 & 0.1\\
\ZllGamma~\cite{PDG2022,ALEPH:2005ab} & 0.5 & 0.1\\
Tau energy scale & 0.4 & 0.1\\
Muon momentum scale & 0.3 & 0.1\\
$W(\rightarrow \ell \nu)+$jets misid. lepton estimate & 0.3 & 0.1\\
(Forward) jet vertex tagging & 0.2 & < 0.1\\
Top subtraction scheme & 0.2 & < 0.1 \\
Electron energy scale & 0.1 & < 0.1 \\
\midrule
Systematic & 12 & 2.4\\
Statistical & 2 & 0.4\\
Total & 13 & 2.5\\
\bottomrule
\end{tabular}
\caption{Impact on \ZinvGamma of different sources of systematic uncertainties, their total, the statistical uncertainty in the data and the overall total uncertainty in \MeV\ and percentage. The individual systematic uncertainties do not sum in quadrature to the total systematic uncertainty due to correlations between uncertainty sources.}\label{tab:Uncertainties}
\end{table}

\section{Invisible width measurement}
\label{sec:results}
 
The value of \ZinvGamma is determined via the ratio defined in Equations~(\ref{eqn:ratio}) and~(\ref{eqn:ratioGamma}). In the \Rmiss\ ratio, both the numerator and denominator are extrapolated to a common phase space and are corrected to account for detector-related effects. The particle-level phase space is defined as:
 
\begin{itemize}
\item at least one jet, where the leading jet must have $\pT \geq 110~\GeV$ and $|\eta | < 2.4$,
\item $\Zpt \ge 130~\GeV$.
\end{itemize}
 
Electrons and muons are defined as before any final-state radiation, the so-called 'Born' level, to ensure consistency between the electron, muon, and neutrino regions. Particle-level jets are found by applying the anti-\kt\ algorithm with radius parameter $R = 0.4$ to final-state particles with decay length $c\tau > 10$~mm, excluding the decay products from the \ZBoson~decay. The \Zpt is defined as the \ZBoson~\pT, calculated from either the two neutrinos or the two charged leptons from the \ZBoson~decay.
 
For each bin, $i$, as a function of \Zpt, the ratios of \Zinvj/\Zeej and \Zinvj/\Zmmj are corrected for detector effects using a bin-wise correction factor, defined as
 
\begin{equation}
U_{i,\ell} = \left({\frac{N_i(\Zinvj)}{N_i(\Zllj)}}\right)^{\text{ detector-level}} \Big{/} \left({\frac{N_i(\Zinvj)}{N_i(\Zllj)}} \right)^{\text{ particle-level}},
\label{eqn:correctionFactor}
\end{equation}
 
where $\ell$ is either an electron or muon and $N_i$ is the number of simulated events in that bin for a given process at either detector-level or particle-level. For the \Zees region, the value of $U$ ranges from 0.3 at low \Zpt to 0.6 at high \Zpt. For the \Zmms region, it ranges from 0.4 to 0.7. These factors are shown as a function of the \Zpt in Figure~\ref{fig:unf} of the appendix. They are dominated by the corrections to the full phase space, which range from 0.5 at low \Zpt to 0.9 at high \Zpt. A bin-wise correction of the ratio is justified as the migrations across bins, which are around 30\%, are similar between the \Zinvj and \Zllj processes and therefore largely cancel. This assumption was checked using a Bayesian unfolding method to correct the \Zinvj, \Zeej and \Zmmj distributions individually and then form the ratio. In addition, the correction factors derived from \Shtte are validated against those determined using \MGfnlo~simulations and found to be consistent.
 
An additional correction is applied to the \Zlls regions to extrapolate to a phase space with no \ZBoson~$m_{\ell \ell}$ criterion and to remove contributions from $\gamma^*$ production, which is only present in charged lepton final states. The correction is derived by comparing \PYTHIA\ simulations where the $\gamma^*$ interference and a \ZBoson~$m_{\ell \ell}$ selection is included to \ZBoson~production without interference and no mass selection. This correction factor is roughly 2\% and is applied to the measured ratio and the predictions. Statistical uncertainties from the MC sample and the theoretical scale uncertainties, as described above, are included as uncertainties in the correction factor. The QCD scale uncertainties are on the sub-percent level.
 
The measured values of \Rmiss{} with electrons and muons final states are shown in Figure~\ref{fig:measured-ratio-inclusive-result}. In the electron channel, the increase in the bin at around 600 \GeV\ is attributed to a statistical fluctuation and is found to diminish with the additional Run 2 data. The data is fully corrected for detector-effects and compared with predictions from \Shtte and \MGfnlo. Both the predictions agree well with the data.
 
\begin{figure*}[h!]
\centering
\includegraphics[width=0.49\textwidth]{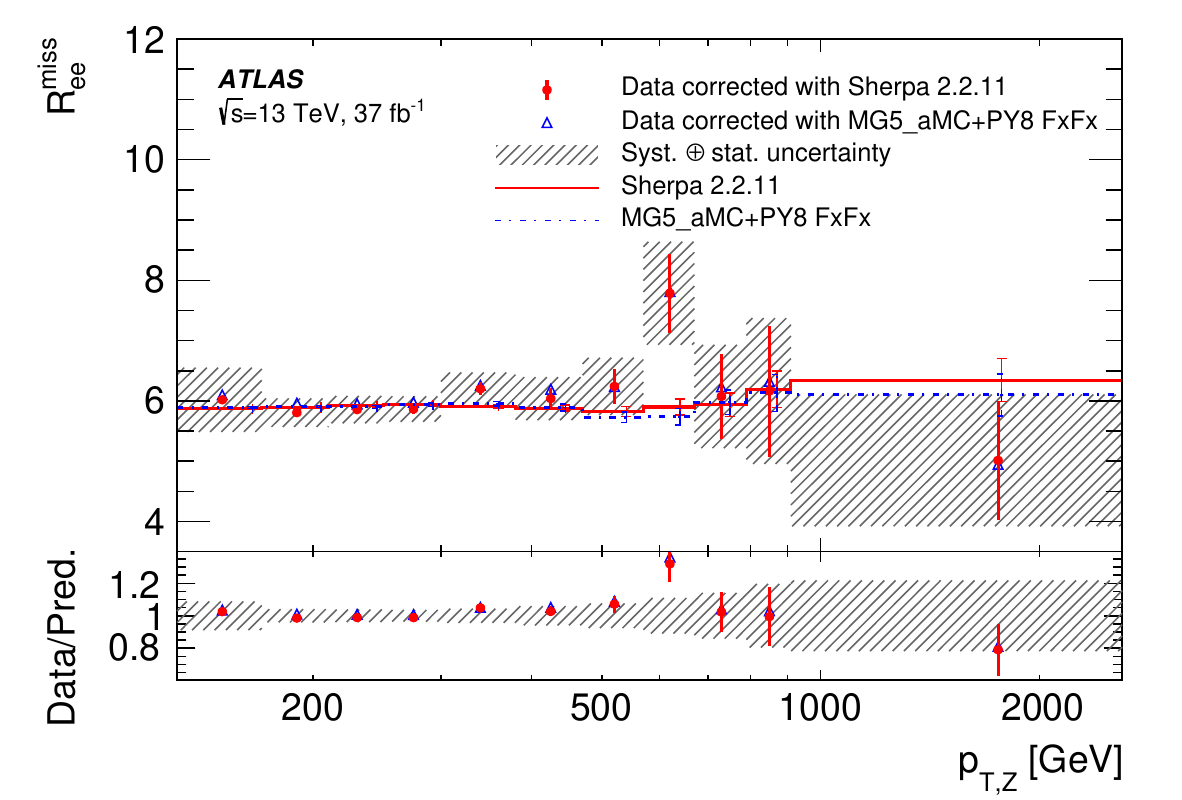}
\includegraphics[width=0.49\textwidth]{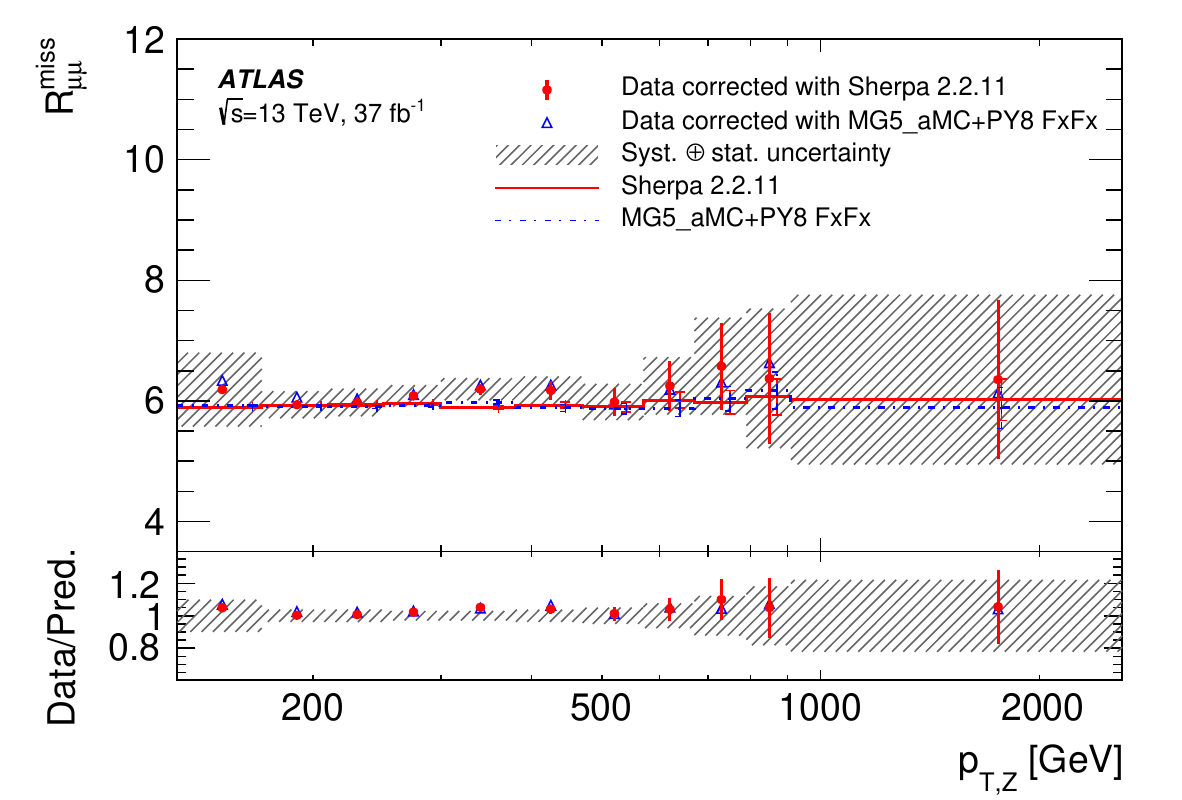}
\caption{Measured \Rmiss{} of (left) electrons and (right) muons as a function of \Zpt in the corrected phase space as defined in Section~\ref{sec:results}. The error bars on the measured red points show the statistical uncertainty and the grey hashed bands show the combined statistical and systematic uncertainties. The results are compared with the prediction for the \Shtte (red) and for the \MGfnlo~(blue) samples. For both the predictions, statistical uncertainties and uncertainties due to the $\gamma^*$ correction are shown by the vertical bars.}
\label{fig:measured-ratio-inclusive-result}
\end{figure*}
 
The value of \Rmisshat\ is determined via a $\chi^2$ minimisation. Determining the value of \Rmisshat\ via a constant fit helps to reduce the systematic uncertainties as the Standard Model predicts that \Rmiss\ should be flat as a function of \Zpt. The value of \ZinvGamma is then determined using Equation~(\ref{eqn:ratioGamma}), utilising the \ZllGamma value, assuming lepton universality, obtained by the LEP experiments~\cite{ALEPH:2005ab, PDG2022}. The function is defined as $\chi^2= (y_{\text{data},I}-\bar{y})^T\mathbf{V}^{-1}(y_{\text{data},I}-\bar{y})$, where the index $I$ runs over all \Zpt bins and both the \Zinv/\Zees and \Zinv/\Zmms ratios, $y_{\text{data},I}$ is the measured \Zinv/\Zees or \Zinv/\Zmms ratio, $\bar{y}$ is the predicted ratio, which is independent of $I$, and $\mathbf{V}^{-1}$ is the inverse covariance matrix.  Statistical and systematic correlations between the \Zinv, \Zees and \Zmms regions are accounted for via the covariance matrix. The lowest \Zpt bin from 130 to 170 \GeV{} is not considered in the minimisation due to large uncertainties in the multijet background.
 
The value of \ZinvGamma obtained is \ZinvValue with a $\chi^2$ value from the fit of 17.3 for 19 degrees of freedom. The values of \ZinvGamma obtained using \Zees and \Zmms individually are \ZinvValueEE and \ZinvValueMM respectively. They are in agreement with each other and with the combined result and also have good $\chi^2$ values. The result is dominated by systematic uncertainties, mainly driven by lepton uncertainties in the denominator of the ratio. The muon channel is more precise and therefore has greater impact in the combination. This is the most precise recoil-based determination of \ZinvGamma. Figure~\ref{fig:resultComp} summarises this result and other recoil-based results. Also shown are the combined value determined from fits of the lineshape of the \Zboson~resonance from LEP, $499.0 \pm 1.5~\MeV$~\cite{ALEPH:2005ab} (assuming lepton universality), and the Standard Model prediction of $501.445 \pm 0.047~\MeV$~\cite{PDG2022}, which is based on inputs such as the Higgs boson and top quark masses. The LEP combination of the photon-tagged results and the result from the lineshape measurements only quote their total uncertainty. The result is compatible with LEP and CMS results and with the Standard Model prediction.

\begin{figure*}[h!]
\centering
\includegraphics[width=0.99\textwidth]{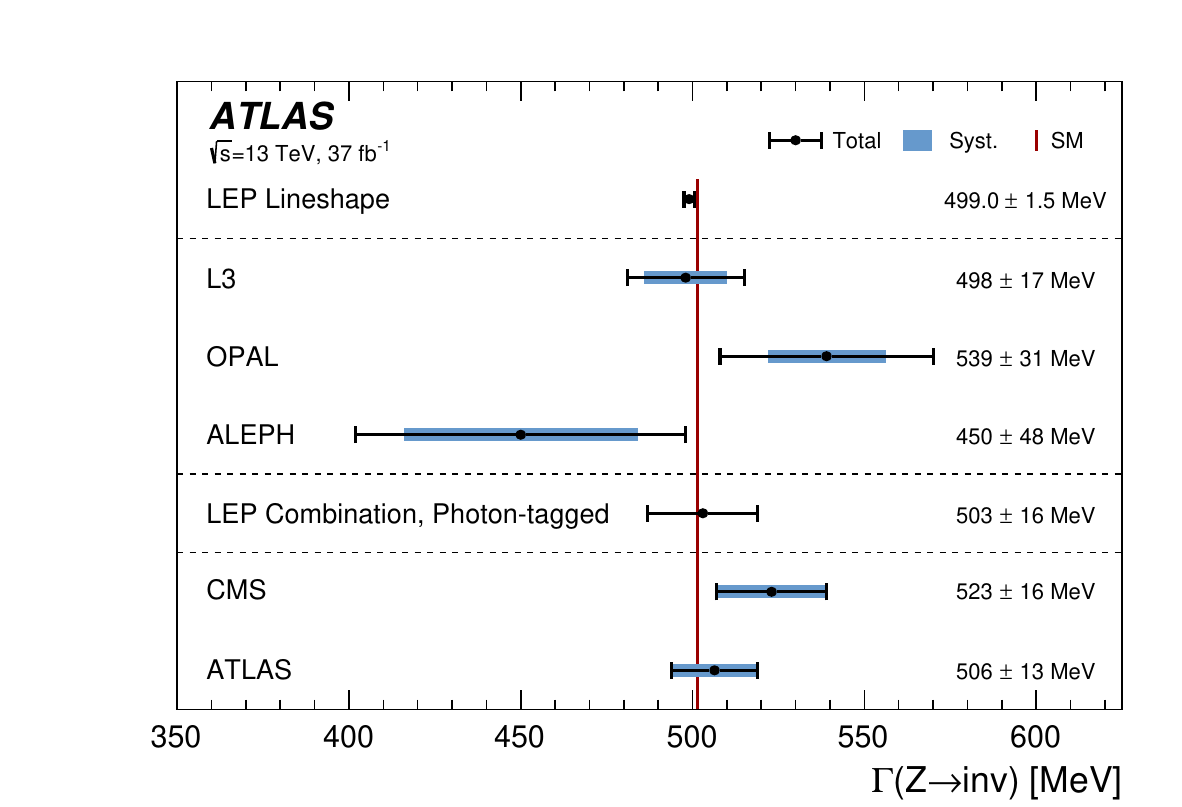}
\caption{\ZinvGamma measured in this paper and by the LEP experiments L3, OPAL, ALEPH and the photon-tagged combination and by the CMS experiment. The total uncertainties are represented by the black error bars and the systematic uncertainty as the blue bands. The LEP combination of the photon-tagged results and the result from the lineshape measurements only quote their total uncertainty. The Standard Model prediction is shown by the solid red line.}
\label{fig:resultComp}
\end{figure*}
 
\FloatBarrier
 
\section{Conclusion}
 
A measurement of the invisible width of the \Zboson boson using 37 \ifb of 13 \TeV\ proton--proton data collected by the ATLAS detector is presented. Measurements of \ZinvGamma in multiple final states are an important consistency test of the Standard Model and thereby a probe of new physics. The ratio of the number of \Zinvj to \Zllj events corrected for detector effects is used. Events with at least one energetic central jet with $\pT \geq 110~\GeV$ are selected for both the \Zinv and \Zlls processes to obtain a similar phase space between the numerator and the denominator of the ratio.
 
The obtained value of \ZinvValue is the most precise experimental result for recoil-based final states to date. The result is in agreement with other recoil-based measurements, with the most precise determination of \ZinvGamma from LEP and with the Standard Model prediction based on three neutrino generations. The dominant uncertainties of this result are due to lepton uncertainties in the denominator of the ratio. A future combination of the recoil-based measurements could further improve the precision of this fundamental parameter of the Standard Model.


We thank CERN for the very successful operation of the LHC and its injectors, as well as the support staff at
CERN and at our institutions worldwide without whom ATLAS could not be operated efficiently.
 
The crucial computing support from all WLCG partners is acknowledged gratefully, in particular from CERN, the ATLAS Tier-1 facilities at TRIUMF/SFU (Canada), NDGF (Denmark, Norway, Sweden), CC-IN2P3 (France), KIT/GridKA (Germany), INFN-CNAF (Italy), NL-T1 (Netherlands), PIC (Spain), RAL (UK) and BNL (USA), the Tier-2 facilities worldwide and large non-WLCG resource providers. Major contributors of computing resources are listed in Ref.~\cite{ATL-SOFT-PUB-2023-001}.
 
We gratefully acknowledge the support of ANPCyT, Argentina; YerPhI, Armenia; ARC, Australia; BMWFW and FWF, Austria; ANAS, Azerbaijan; CNPq and FAPESP, Brazil; NSERC, NRC and CFI, Canada; CERN; ANID, Chile; CAS, MOST and NSFC, China; Minciencias, Colombia; MEYS CR, Czech Republic; DNRF and DNSRC, Denmark; IN2P3-CNRS and CEA-DRF/IRFU, France; SRNSFG, Georgia; BMBF, HGF and MPG, Germany; GSRI, Greece; RGC and Hong Kong SAR, China; ISF and Benoziyo Center, Israel; INFN, Italy; MEXT and JSPS, Japan; CNRST, Morocco; NWO, Netherlands; RCN, Norway; MEiN, Poland; FCT, Portugal; MNE/IFA, Romania; MESTD, Serbia; MSSR, Slovakia; ARRS and MIZ\v{S}, Slovenia; DSI/NRF, South Africa; MICINN, Spain; SRC and Wallenberg Foundation, Sweden; SERI, SNSF and Cantons of Bern and Geneva, Switzerland; MOST, Taipei; TENMAK, T\"urkiye; STFC, United Kingdom; DOE and NSF, United States of America.
 
Individual groups and members have received support from BCKDF, CANARIE, CRC and DRAC, Canada; PRIMUS 21/SCI/017 and UNCE SCI/013, Czech Republic; COST, ERC, ERDF, Horizon 2020, ICSC-NextGenerationEU and Marie Sk{\l}odowska-Curie Actions, European Union; Investissements d'Avenir Labex, Investissements d'Avenir Idex and ANR, France; DFG and AvH Foundation, Germany; Herakleitos, Thales and Aristeia programmes co-financed by EU-ESF and the Greek NSRF, Greece; BSF-NSF and MINERVA, Israel; Norwegian Financial Mechanism 2014-2021, Norway; NCN and NAWA, Poland; La Caixa Banking Foundation, CERCA Programme Generalitat de Catalunya and PROMETEO and GenT Programmes Generalitat Valenciana, Spain; G\"{o}ran Gustafssons Stiftelse, Sweden; The Royal Society and Leverhulme Trust, United Kingdom.
 
In addition, individual members wish to acknowledge support from Chile: Agencia Nacional de Investigación y Desarrollo (FONDECYT 1190886, FONDECYT 1210400, FONDECYT 1230987); China: National Natural Science Foundation of China (NSFC - 12175119, NSFC 12275265); European Union: European Research Council (ERC - 948254), Horizon 2020 Framework Programme (MUCCA - CHIST-ERA-19-XAI-00), Italian Center for High Performance Computing, Big Data and Quantum Computing (ICSC, NextGenerationEU), Marie Sklodowska-Curie Actions (EU H2020 MSC IF GRANT NO 101033496); France: Agence Nationale de la Recherche (ANR-20-CE31-0013, ANR-21-CE31-0022), Investissements d'Avenir Idex (ANR-11-LABX-0012), Investissements d'Avenir Labex (ANR-11-LABX-0012); Germany: Baden-Württemberg Stiftung (BW Stiftung-Postdoc Eliteprogramme), Deutsche Forschungsgemeinschaft (DFG - CR 312/5-1); Italy: Istituto Nazionale di Fisica Nucleare (FELLINI G.A. n. 754496, ICSC, NextGenerationEU); Japan: Japan Society for the Promotion of Science (JSPS KAKENHI 22H01227, JSPS KAKENHI JP21H05085, JSPS KAKENHI JP22H04944); Netherlands: Netherlands Organisation for Scientific Research (NWO Veni 2020 - VI.Veni.202.179); Norway: Research Council of Norway (RCN-314472); Poland: Polish National Agency for Academic Exchange (PPN/PPO/2020/1/00002/U/00001), Polish National Science Centre (NCN 2021/42/E/ST2/00350, NCN OPUS nr 2022/47/B/ST2/03059, NCN UMO-2019/34/E/ST2/00393, UMO-2020/37/B/ST2/01043, UMO-2021/40/C/ST2/00187); Slovenia: Slovenian Research Agency (ARIS grant J1-3010); Spain: BBVA Foundation (LEO22-1-603), Generalitat Valenciana (Artemisa, FEDER, IDIFEDER/2018/048), La Caixa Banking Foundation (LCF/BQ/PI20/11760025), Ministry of Science and Innovation (RYC2019-028510-I, RYC2020-030254-I), PROMETEO and GenT Programmes Generalitat Valenciana (CIDEGENT/2019/023, CIDEGENT/2019/027); Sweden: Swedish Research Council (VR 2022-03845); Switzerland: Swiss National Science Foundation (SNSF - PCEFP2\_194658); United Kingdom: Leverhulme Trust (Leverhulme Trust RPG-2020-004); United States of America: Neubauer Family Foundation.


\clearpage
\appendix
\part*{Appendix}
\addcontentsline{toc}{part}{Appendix}
 
A breakdown of the systematic uncertainties in the \Zinv, \Zees and \Zmms\ regions is shown in Figure~\ref{fig:RecoUnc}. The systematic uncertainties in \Rmiss{}, after all corrections are applied, are indicated in Figure~\ref{fig:unfUnc}. The values of the correction factor, $U_{i,\ell}$ are shown in Figure~\ref{fig:unf} as a function of \Zpt for both the \Zees and \Zmms regions.
 
\begin{figure*}[h!]
\centering
\includegraphics[width=0.51\textwidth]{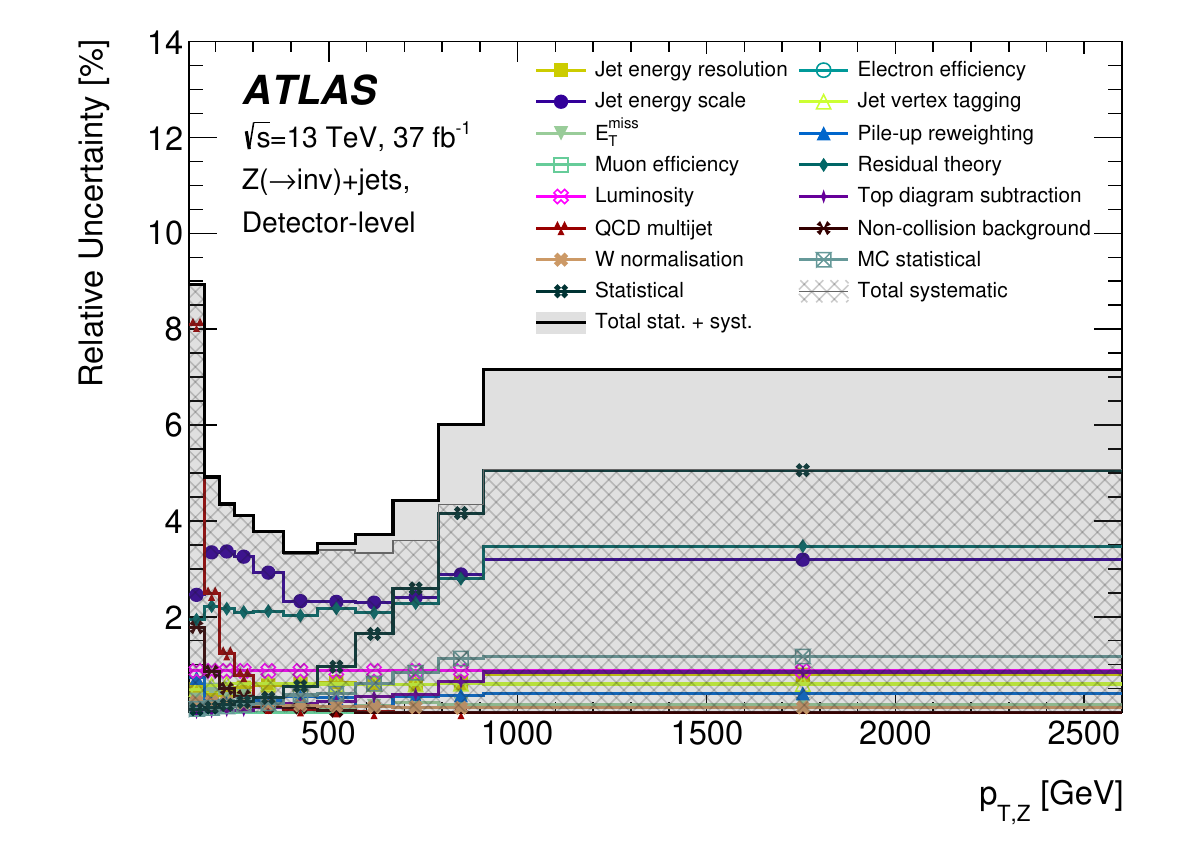}
\includegraphics[width=0.49\textwidth]{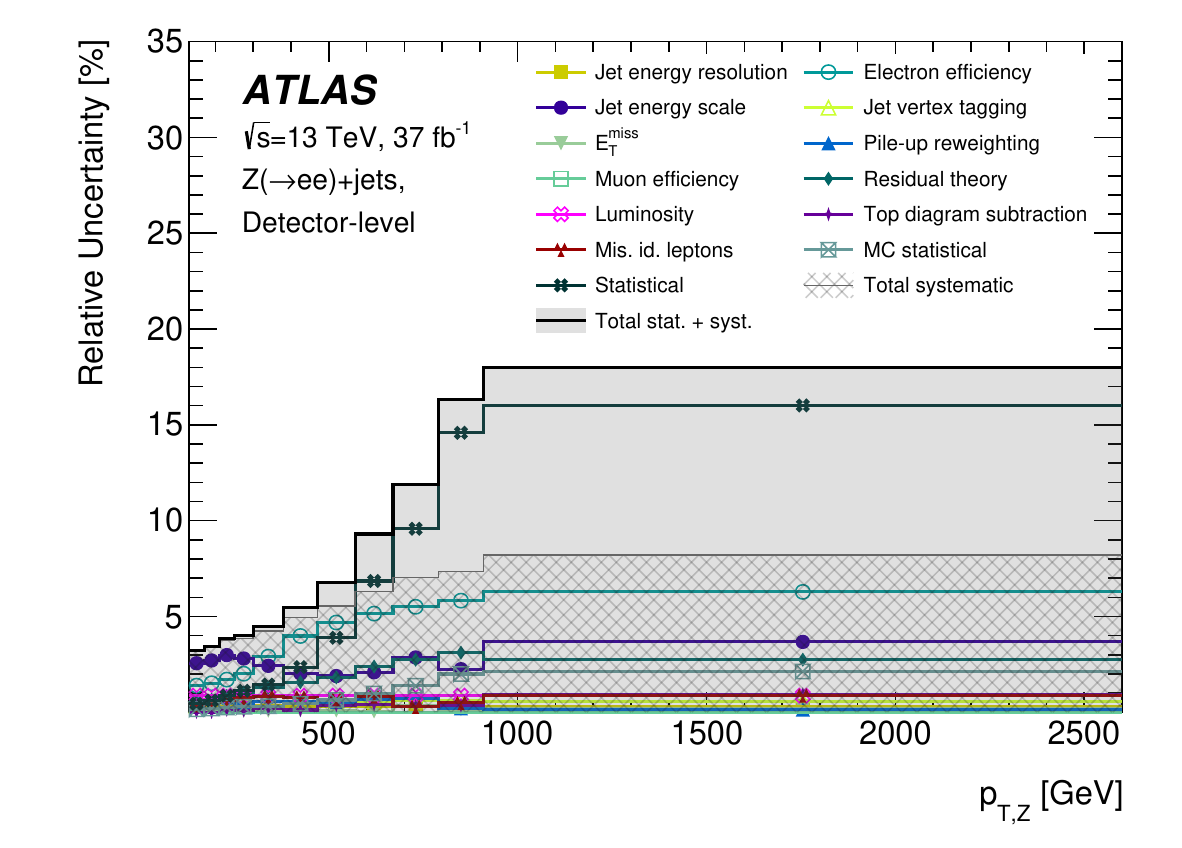}
\includegraphics[width=0.49\textwidth]{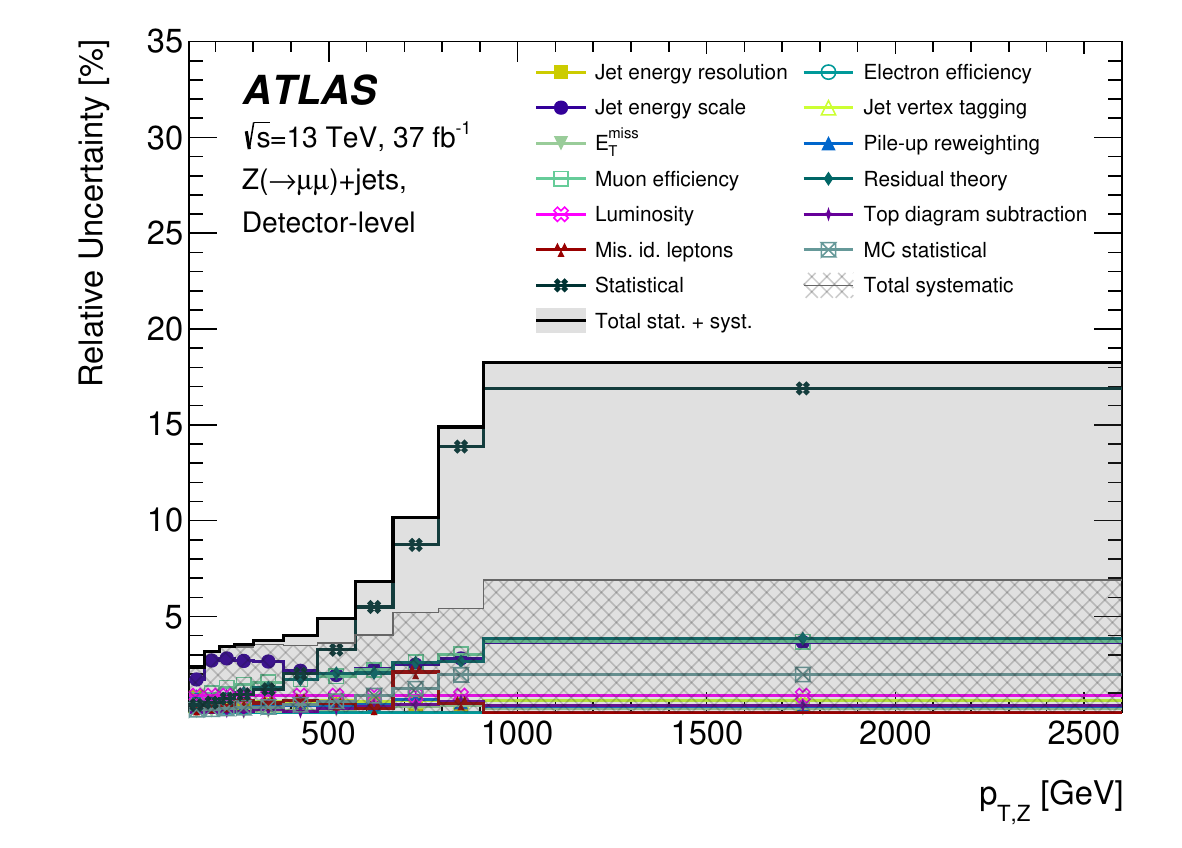}
\caption{Relative uncertainties in percentage in the number of predicted events, as a function of \Zpt for \Zinv (top), \Zees (bottom left) and \Zmms (bottom right) regions before detector corrections. Only uncertainties that are larger than 0.5\% in any bin are shown.}
\label{fig:RecoUnc}
\end{figure*}
 
\begin{figure*}[h!]
\centering
\includegraphics[width=0.49\textwidth]{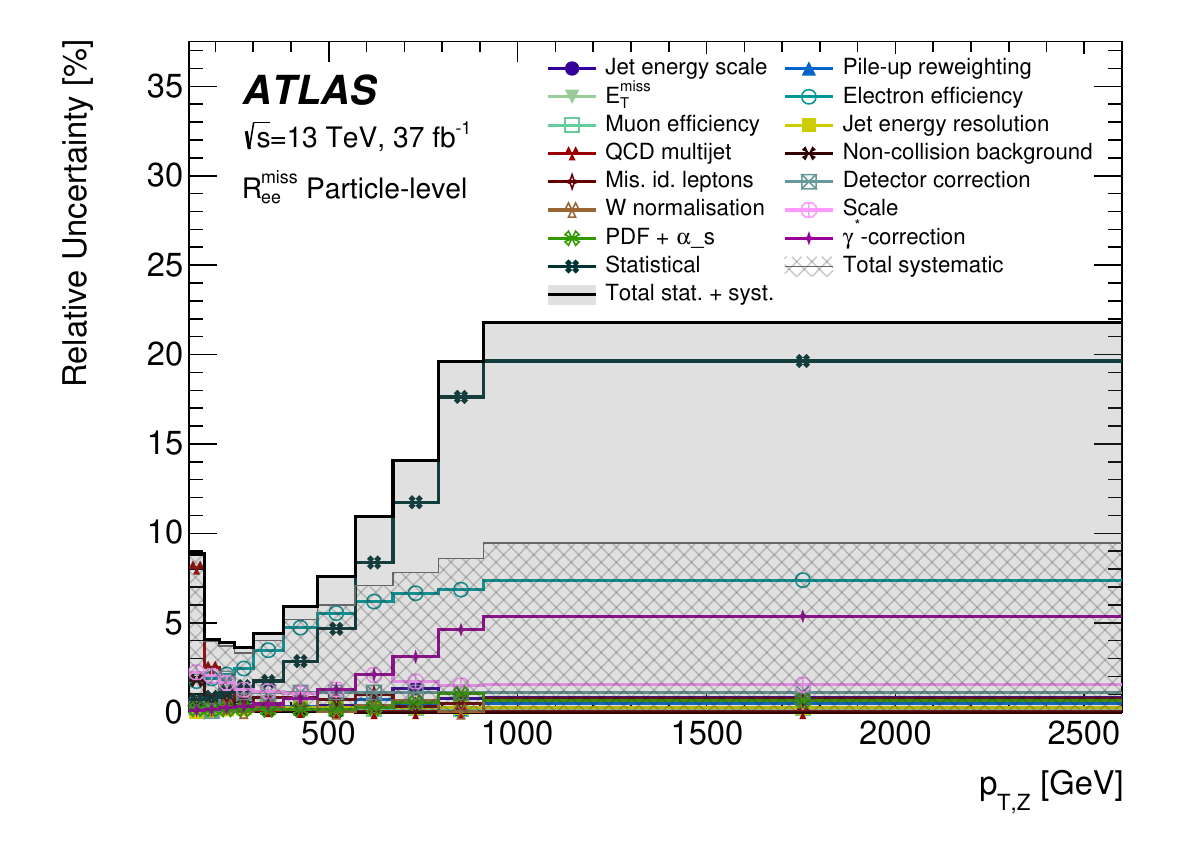}
\includegraphics[width=0.49\textwidth]{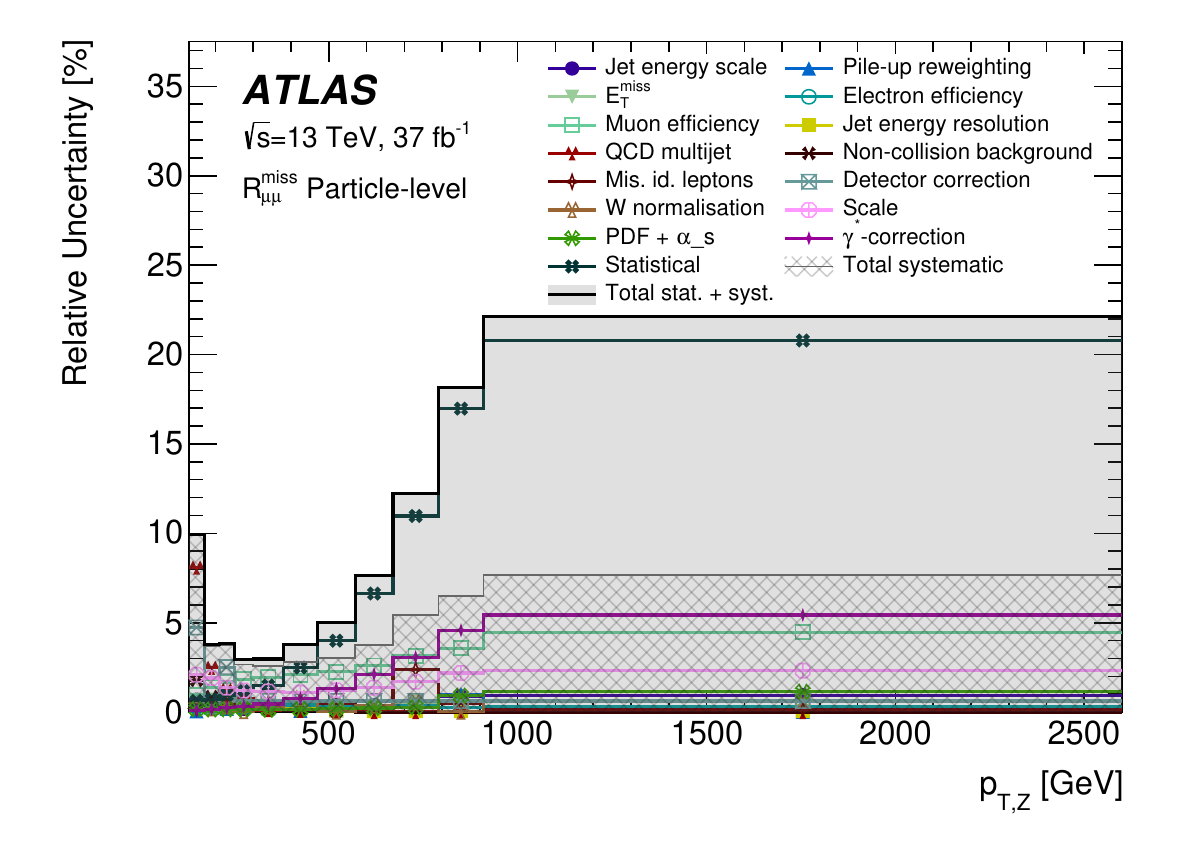}
\caption{Relative uncertainties in percentage in \Rmiss{}, as a function of \Zpt for the (left) electron and (right) muon channels after detector corrections. Only uncertainties that are larger than 0.5\% in any bin are shown.}
\label{fig:unfUnc}
\end{figure*}

\begin{figure*}[h!]
\centering
\includegraphics[width=0.51\textwidth]{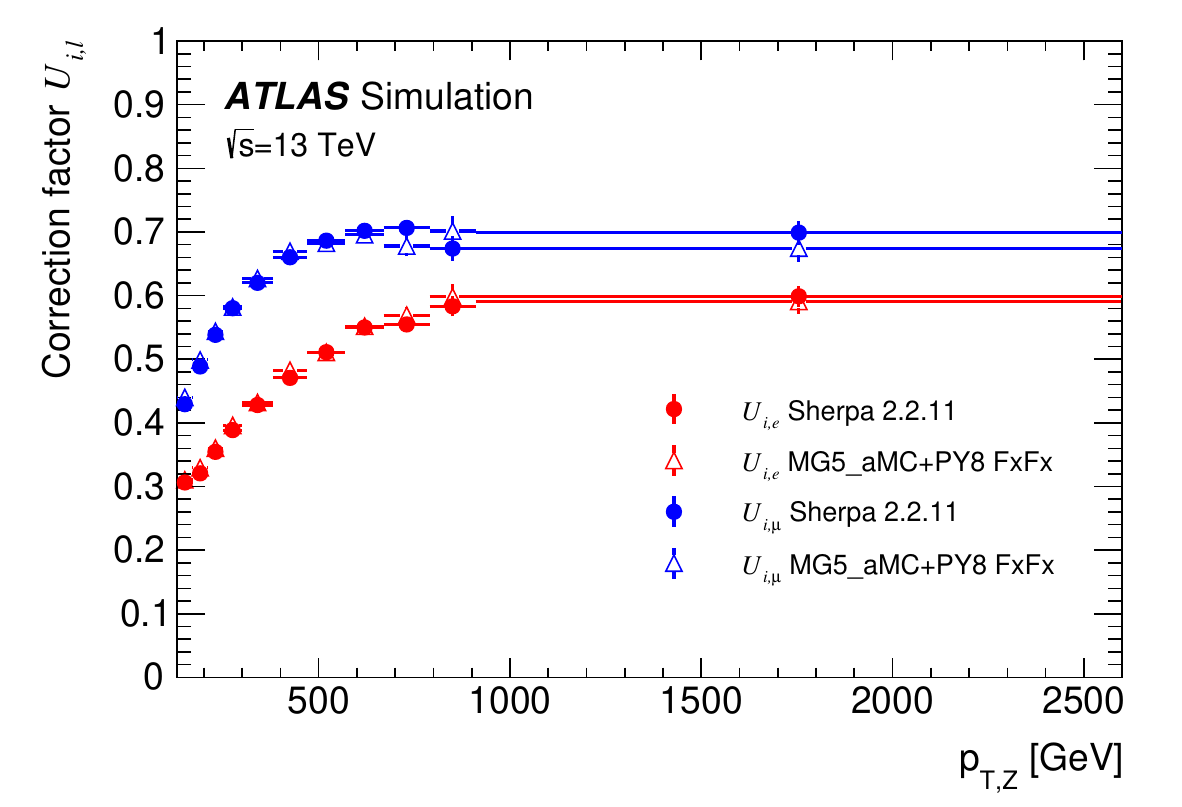}
\caption{The value of the correction factor, $U_{i,\ell}$, as a function of \Zpt for both the \Zees (in red) and \Zmms (in blue) regions. Shown in solid dots is the correction as determined with \Shtte~simulations and in open triangles that determined with \MGfnlo~simulations. Only the Monte Carlo statistical uncertainties are shown.}
\label{fig:unf}
\end{figure*}
 
\clearpage

\printbibliography
 
\clearpage
\input{atlas_authlist}

\end{document}

%% file: atlas_authlist.tex
 
\begin{flushleft}
\hypersetup{urlcolor=black}
{\Large The ATLAS Collaboration}

\bigskip

\AtlasOrcid[0000-0002-6665-4934]{G.~Aad}$^\textrm{\scriptsize 102}$,
\AtlasOrcid[0000-0002-5888-2734]{B.~Abbott}$^\textrm{\scriptsize 120}$,
\AtlasOrcid[0000-0002-1002-1652]{K.~Abeling}$^\textrm{\scriptsize 55}$,
\AtlasOrcid[0000-0001-5763-2760]{N.J.~Abicht}$^\textrm{\scriptsize 49}$,
\AtlasOrcid[0000-0002-8496-9294]{S.H.~Abidi}$^\textrm{\scriptsize 29}$,
\AtlasOrcid[0000-0002-9987-2292]{A.~Aboulhorma}$^\textrm{\scriptsize 35e}$,
\AtlasOrcid[0000-0001-5329-6640]{H.~Abramowicz}$^\textrm{\scriptsize 151}$,
\AtlasOrcid[0000-0002-1599-2896]{H.~Abreu}$^\textrm{\scriptsize 150}$,
\AtlasOrcid[0000-0003-0403-3697]{Y.~Abulaiti}$^\textrm{\scriptsize 117}$,
\AtlasOrcid[0000-0002-8588-9157]{B.S.~Acharya}$^\textrm{\scriptsize 69a,69b,m}$
% ...
\bigskip
\\

% ...
$^{29}$Physics Department, Brookhaven National Laboratory, Upton NY; United States of America.\\
% ...
$^{35}$$^{(a)}$Facult\'e des Sciences Ain Chock, R\'eseau Universitaire de Physique des Hautes Energies - Universit\'e Hassan II, Casablanca;$^{(b)}$Facult\'{e} des Sciences, Universit\'{e} Ibn-Tofail, K\'{e}nitra;$^{(c)}$Facult\'e des Sciences Semlalia, Universit\'e Cadi Ayyad, LPHEA-Marrakech;$^{(d)}$LPMR, Facult\'e des Sciences, Universit\'e Mohamed Premier, Oujda;$^{(e)}$Facult\'e des sciences, Universit\'e Mohammed V, Rabat;$^{(f)}$Institute of Applied Physics, Mohammed VI Polytechnic University, Ben Guerir; Morocco.\\
% ...
$^{49}$Fakult\"{a}t Physik , Technische Universit{\"a}t Dortmund, Dortmund; Germany.\\
% ...
$^{55}$II. Physikalisches Institut, Georg-August-Universit\"{a}t G\"ottingen, G\"ottingen; Germany.\\
% ...
$^{69}$$^{(a)}$INFN Gruppo Collegato di Udine, Sezione di Trieste, Udine;$^{(b)}$ICTP, Trieste;$^{(c)}$Dipartimento Politecnico di Ingegneria e Architettura, Universit\`a di Udine, Udine; Italy.\\
% ...
$^{102}$CPPM, Aix-Marseille Universit\'e, CNRS/IN2P3, Marseille; France.\\
% ...
$^{117}$Department of Physics, New York University, New York NY; United States of America.\\
% ...
$^{120}$Homer L. Dodge Department of Physics and Astronomy, University of Oklahoma, Norman OK; United States of America.\\
% ...
$^{150}$Department of Physics, Technion, Israel Institute of Technology, Haifa; Israel.\\
$^{151}$Raymond and Beverly Sackler School of Physics and Astronomy, Tel Aviv University, Tel Aviv; Israel.\\
% ...
$^{a}$ Also Affiliated with an institute covered by a cooperation agreement with CERN.\\
$^{b}$ Also at An-Najah National University, Nablus; Palestine.\\
$^{c}$ Also at Borough of Manhattan Community College, City University of New York, New York NY; United States of America.\\
$^{d}$ Also at Center for High Energy Physics, Peking University; China.\\
$^{e}$ Also at Center for Interdisciplinary Research and Innovation (CIRI-AUTH), Thessaloniki; Greece.\\
$^{f}$ Also at Centro Studi e Ricerche Enrico Fermi; Italy.\\
% ...
$^{m}$ Also at Department of Physics, King's College London, London; United Kingdom.\\
% ...

\end{flushleft}
